\documentstyle[12pt,epsfig]{article}

\def\l{\lambda}
\def\beq{\begin{equation}}
\def\eeq{\end{equation}}

\begin{document}

\begin{titlepage}
 \hfill       OHSTPY-HEP-T-00-013\\
  \mbox{ } \hfill      August 29, 2000 \\

\vspace{1.0cm}
    
\begin{center}

  {\Large\bf A solution to the $\mu$ problem}

\vspace{.5cm}
  
  {\Large\bf in the presence of a heavy gluino LSP}

  \vspace{2cm}
  
  {\large\bf Arash Mafi and Stuart Raby}
    
    \bigskip  
{\em Department of Physics, The Ohio State University, 174 W. 18th Ave.,
Columbus, Ohio  43210}
      
  \vspace{1cm}
{\bf Abstract}
\end{center}

In this paper we present a solution to the $\mu$ problem in an
SO(10) supersymmetric grand unified model with gauge mediated and
D-term supersymmetry breaking.   A Peccei-Quinn
symmetry is broken at the messenger scale $M\sim 10^{12}$ GeV and enables
the generation of the $\mu$ term. The boundary conditions
defined at $M$ lead to a phenomenologically
acceptable version of the minimal supersymmetric standard model with
novel particle phenomenology.
Either the gluino or the gravitino is the lightest supersymmetric particle (LSP). 
If the gravitino is the LSP, then the gluino is the next-to-LSP (NLSP) with 
a lifetime on the order of one month or longer. In either case this heavy
gluino, with mass in the range 25 - 35 GeV, can be treated as a stable particle 
with respect to experiments at high energy accelerators.  Given the extensive 
phenomenological constraints we show that the model can only survive
in a narrow region of parameter space resulting in a light neutral Higgs with 
mass $\sim 86 - 91$ GeV and $\tan\beta \sim 9 - 14$.  In addition the lightest 
stop and neutralino have mass $\sim 100 - 122$ GeV and $\sim 50 - 72$ GeV,
respectively.  Thus the model will soon be tested.  Finally, the invisible axion 
resulting from PQ symmetry breaking is a cold dark matter candidate.

\end{titlepage}

\section{Introduction}
\label{sec:intro}

Supersymmetry (SUSY) is a strongly motivated candidate for new physics 
beyond the Standard Model (SM). It provides a natural framework for resolving the 
hierarchy problem.  The minimal supersymmetric standard model (MSSM), with a conserved 
R-parity, has an economical particle content with well defined interactions most of 
which are already constrained by experiment.  It has two Higgs doublets ($H_u$ and $H_d$); 
necessary for giving mass to both up and down quarks, respectively.  In the MSSM, electroweak 
symmetry breaking (EWSB) occurs naturally since $m^2_{H_u}$, the soft SUSY breaking (SSB) 
mass of $H_u$, is automatically driven negative as a result of a large top quark Yukawa coupling. 

The MSSM solves the hierarchy problem by allowing for dimensionful SSB parameters of order the 
electroweak scale and protecting scalar masses from large radiative corrections above the SSB 
scale.  However, this in itself is not sufficient to solve the hierarchy problem.   In addition, 
it is necessary to demand that the $\mu$ parameter (where $\mu$ is the bilinear Higgs coupling 
in the superpotential of the form $\mu H_u H_d$) is also of order the electroweak scale.  
Consider the vacuum conditions obtained by minimizing the tree level Higgs potential, we have
\beq
\label{eq:musquared}
\mu^2=-\frac{M_Z^2}{2}+
\frac{m^2_{H_d}- m^2_{H_u}tan^2\beta}{tan^2\beta-1},
\eeq
where $\tan{\beta}=\langle H_u\rangle/ \langle H_d\rangle$, $sin 2\beta = 2 B/m_A^2$, $m_A$ 
is the mass of the CP odd Higgs
and $B$ is the SSB Higgs bilinear coupling.
On the left hand side of Eq. \ref{eq:musquared}, the $\mu$ parameter,
multiplying a supersymmetric $\mu$ term in the Lagrangian, breaks no 
SM symmetries; it could in principle be as large as the Planck or GUT scales.  
On the right side, the $Z$ boson mass and the SSB Higgs masses are of order the 
electroweak scale.  Clearly, all three scales, i.e.  $\mu$, $M_Z$ and the
SSB scales must be of the same order. This is the $\mu$ problem \cite{MuNir}.

In order to avoid large values for $\mu$, a symmetry is needed which prevents the $\mu$ term at tree level, but allows such a term once this symmetry is broken.  Moreover, since in the MSSM the $\mu$ term is contained in the superspace potential,  there are two possibilities - 1) it can be generated via a term in the Kahler potential (at tree level or radiatively) once supersymmetry is broken or 2) no supersymmetry breaking is required if it is generated through higher dimension operators in the superspace potential.
Several simple mechanisms for generating a $\mu$ term have been suggested
\cite{KimNilles,GiudiceMasiero}.  

In the context of gauge mediated supersymmetry breaking (GMSB) 
models \cite{gmsb}, there is an additional problem.  $\mu$ can be generated at one loop order once supersymmetry is broken. However,
the $B$ parameter (the SSB scalar Higgs bilinear coupling)
is usually generated at the same loop order and is too large.
A solution generating $B$ at higher loop order than $\mu$ was proposed in \cite{DvGiPo}.

In this paper we use an extension of the GMSB model discussed
in Ref.\ \cite{RabyTobe3} to solve the $\mu$ problem. 
This model has GMSB with Higgs-messenger mixing in an SO(10) theory
and naturally leads to a gluino LSP.   The gluino LSP is stable due to 
R-parity conservation.  The specific signature of a gluino LSP i.e. missing momentum has been analysed in Ref.\ \cite{BCG} for LEP and CDF data 
and in Ref.\ \cite{RabyMafi} for CDF data.  Ref.\ \cite{RabyMafi} concludes that a stable gluino with mass in the range $25 - 35$ GeV is still allowed by 
both the LEP and CDF data.  Our model, with a modest adjustment of parameters,
gives a gluino with mass in this range.

The $\mu$ term is absent in this model (at tree level) due to a U(1) Peccei-Quinn ({\bf PQ}) symmetry.  Both SUSY and {\bf PQ} symmetry are broken when the chiral superfield $X$ develops a vacuum expectation value (vev)
\beq
\langle X\rangle=M+\theta^2 F_X .
\eeq
Note, the following dimension five operator in the Kahler potential
\beq
K \supset \frac{1 }{M_P} X^\dagger {\bf 10}_H^2 + h.c.
\eeq
is however allowed by the symmetries.  Thus we find $\mu \sim F_X /M_P$. 
$B$, on the other hand, is generated radiatively via renormalization group (RG)
running below the messenger scale $M$.\footnote{Note, the mechanism used here to generate the $\mu$ term is a combination of the ideas discussed in \cite{KimNilles,GiudiceMasiero}.}

In addition to solving the $\mu$ problem, the {\bf PQ} symmetry 
provides a natural solution to the strong CP problem \cite{PecQui}. 
The strong CP violating $\theta$ term dynamically tracks to zero.
Moreover, as a bonus, the axion is a candidate for cold dark matter.  

In the next section we discuss the model, saving some of the details for
the appendices.   We derive the low energy spectrum consistent with
electroweak symmetry breaking, gauge coupling unification and third generation quark and lepton masses.  We then consider experimental constraints which
constrain the available parameter space to a very narrow region.
In this region we find a light neutral Higgs with mass $\sim 86 - 91$ GeV and $\tan\beta \sim 9 - 14$.  In addition the lightest stop and neutralino have mass $\sim 100 - 122$ GeV and $\sim 50 - 72$ GeV, respectively.  Finally, in 
an appendix we investigate the possibility of obtaining a reasonable mass
for the tau neutrino in the model.  Clearly this model is preeminently testable.   

\section{The model}
\label{sec:model}

The theory at the GUT scale is defined by the SO(10)
invariant superpotential $W\supset W_1+W_2+W_3$ 
and a non-renormalizable term in the Kahler
potential $K$ where

\begin{eqnarray}
\begin{array}{c}
W_1={\bf 16}_3 {\bf 10}_H {\bf 16}_3,\\ \\
W_2=\l_a{\bf 10}_H A {\bf 10}_A+\l_X X {\bf 10}_A^2,\\ \\
W_3=\l_1 \bar{\eta}_1 A\eta_1+
\l_2\bar{\eta}_2 A {\eta}_2+\l X\bar{\bf \eta}_1{\bf \eta}_2.
\end{array}
\end{eqnarray}

\beq
K\supset \l_K\frac{X^\dagger}{M_P}{\bf 10}_H^2+h.c.
\eeq
$({\bf 16}_3,\; \eta_1,\; \eta_2)$ are ${\bf 16}$'s,
$(\bar{\eta}_1,\; \bar{\eta}_2)$ are
$\bar{\bf 16}$'s, 
$({\bf 10}_H,\; {\bf 10}_A)$ are ${\bf 10}$'s,
$(X)$ is a singlet and $(A)$ is an adjoint under SO(10).  

At the GUT scale, the theory is invariant 
under a U(1) {\bf PQ} and an R symmetry.
The R symmetry is broken spontaneously at the GUT scale. 
The {\bf PQ} symmetry, however, is not broken at the GUT scale and prevents a $\mu$ term in the superpotential.  The {\bf PQ} and R charges of the fields are
defined in Appendix \ref{sec:completemodel}.

$W_1$ contains the coupling of the third family matter multiplet (${\bf 16}_3$) 
to the Higgs field  (${\bf 10}_H$) which includes  both the weak doublet and color triplet Higgs fields.  

$W_2$ serves two purposes.  In the first case, it provides doublet-triplet splitting using the Dimopoulos-Wilczek mechanism \cite{DimWil}.    
The adjoint field $A$ gets a vev
\beq
\label{eq:vevA}
\langle A \rangle=(B-L)M_G,
\eeq
where $B-L$ (baryon number minus lepton number) is non-vanishing
on color triplets and zero on weak doublets and the 
singlet $X$ gets a vev 
\beq
\langle X\rangle=M+\theta^2 F_X.
\eeq
This gives mass of order $M_G$ to the color triplet Higgs states and of order
$M$ to the weak doublets in ${\bf 10}_A$.   The Higgs doublets in ${\bf 10}_H$
remain massless.   The SUSY breaking vev, on the other hand, exhibits the second
purpose for $W_2$.

In the second case, $W_2$ and $W_3$ also provide the messengers for SUSY breaking.\footnote{Due to an accidental cancellation, gluinos receive no mass at one loop from $W_2$.  Thus $W_3$ is introduced with additional messenger fields $(\eta_1,\bar{\eta}_1,\eta_2,\bar{\eta}_2)$ contributing to the masses of gauginos and scalars at the scale $M_G$.}
   The auxiliary field ${\bf 10}_A$ and the fields $\bar \eta_1, \eta_2$ feel SUSY breaking at tree level due to the vev $F_X$.   They are thus
the messengers for GMSB \cite{RabyTobe3, Raby1}.
We take the messenger scale $M\sim 10^{12}$ GeV with the effective SUSY breaking scale in the observable sector given by
\beq
\Lambda=F_X/M\sim 10^5\ GeV.
\eeq

Thus the Higgs field in this model plays a central role with regards to supersymmetry breaking.  It is this central role which also provides a
natural framework for solving the $\mu$ problem using the
{\bf PQ} symmetry.  When $X$ gets a vev, both SUSY and the {\bf PQ} symmetry are broken.   The $\mu$ term is generated at the scale $M$
\beq
\mu=\l_K\frac{F_X}{M_P}.
\eeq
while $B$ remains zero at tree level. 

The {\bf PQ} symmetry solves the strong CP problem and produces an axion;
the Goldstone boson of the broken {\bf PQ} symmetry. The axion gets mass due 
to the QCD chiral anomaly  of order $m^2_a = \frac{f_\pi^2}{f_a^2} m_\pi^2 N^2 \frac{Z}{(1+Z)^2}$ \cite{DFS}   where $Z = m_u/m_d \sim 0.56$, $f_a \equiv M = 10^{12}$ GeV is the {\bf PQ} symmetry breaking scale and $N=3$ is the number of families.  Putting in the numbers we find  $m_a \sim 2\times 10^{-5}$ eV.\footnote{We take $M = 10^{12}$ GeV, so that the energy density in the axion field does not over close the universe.}

We refer the interested reader to Appendix \ref{sec:completemodel} for
the complete model defined at the GUT scale.  Note, in order to obtain realistic $t,\;  b$ and $\tau$ masses, we find it necessary to abandon Yukawa unification at $M_G$.  How this is obtained in the complete model is discussed in Appendix \ref{sec:nonunif}.  Finally, a simple extension of the model to include a
$\tau$ neutrino mass is presented in Appendix \ref{sec:neutrinomass}.

\section{Boundary conditions at the messenger scale}
\label{sec:messenger}

The boundary conditions at the messenger scale are determined
by two sources of SUSY breaking, gauge mediation and D-term \cite{RabyTobe3}.
The messengers give mass to the gauginos and Higgs at one loop and to squarks and sleptons at two loops.  Since the color triplet messengers have mass of order the GUT scale, the gluino mass is suppressed compared to the other
gauginos.  The gaugino masses (at $M$) are given by

\begin{eqnarray}
\label{eq:gauginomass}
\begin{array}{cc}
m_{\tilde{g}}=\frac{\alpha_3}{\pi}\Lambda b^2,\\ \\

M_2=\frac{\alpha_2}{4\pi} \Lambda(1+\frac{28}{9} b^2),\\ \\

M_1=\frac{3}{5}\frac{\alpha_1}{4\pi} \Lambda(1+4 b^2).

\end{array}
\end{eqnarray}
where
\beq
b^2=-\frac{9\l^2}{\l_1\l_2} \frac{M^2}{M_G^2} > 0 .
\eeq
The two loop GMSB contribution to the scalar masses is given by
\beq
\label{eq:scalarmass}
\tilde{m}^2=2\Lambda^2
\{C_3(\frac{\alpha_3}{4\pi})^2(a^2+4b^2)+
C_2(\frac{\alpha_2}{4\pi})^2(1+\frac{28}{9}b^2)+
C_1(\frac{\alpha_1}{4\pi})^2(\frac{3}{5}+
{2\over 5}a^2+{12\over 5}b^2)\},
\eeq
where $a=\frac{\l_X M}{\l_a M_G}$, $C_3=4/3$ for color triplets
and zero otherwise, $C_2=3/4$ for weak doublets and zero
otherwise and $C_1=(\frac{3}{5})(Y/2)^2$. $a$ and $b$ are free 
independent parameters which we use to fit the data at
the EWSB scale. 

The gravitino mass is given by  
\beq
\label{eq:gravitinomass} 
m_{\tilde{G}}=\frac{\Lambda_{susy}^2}{\sqrt{3} M_P}, 
\eeq
where $\Lambda_{susy}$ is the scale of SUSY breaking and
$M_P=(8\pi G_N)^{-1/2}=2.4\times 10^{18}$ GeV is the reduced Planck
mass. For $\Lambda_{susy}^2=F_X\simeq 10^{17}\ GeV^2$ the gravitino mass is 
\beq
\label{eq:lightgravitino}
m_{\tilde{G}} = \frac{M}{\sqrt{3} M_P} \Lambda \simeq 0.024\ GeV,
\eeq
making the gravitino the LSP.   However, this conclusion is model dependent.

For example, we show that in the particular model of SUSY breaking discussed in
Ref. \cite{RabyTobe4}, the gravitino mass is significantly larger.
In this model the field which gets both a scalar
and F-component vev is the third component of an $SU(2)_F$
vector field $S_3$. In this theory $X$ is a composite field
with $\langle X\rangle=M=S_3^2/M_{st}$ and
$F_X=2S_3F_{S_3}/M_{st}$.  The gravitino mass is therefore given by
\cite{Raby1}
\beq
m_{\tilde{G}}=\frac{F_{S_3}}{\sqrt{3}M_P}=\frac{1}{2\sqrt{3}}
\frac{\sqrt{M M_{st}}}{M_P} \Lambda,
\eeq
with $\Lambda$ still given by $\Lambda=F_X/M$. The gravitino mass is thus
enhanced by the factor  $\frac{1}{2} \sqrt{\frac{M_{st}}{M}}$.  For example, letting the string scale $M_{st}=M_P$ and requiring the scale of {\bf PQ} symmetry breaking $M=10^{12}$ GeV, we find  $ m_{\tilde{G}} = 18.6$ GeV.\footnote{It is necessary to check that the supergravity contribution to squark and slepton masses is small compared to the GMSB contribution.  This ratio scales as
\beq
\frac{m_{\tilde{G}}}{M_2}=\frac{2\pi}{\sqrt{3}\alpha_2}
\frac{S_3}{M_P}.
\eeq
Thus $S_3$ cannot be much larger than $10^{15}$ giving  
$\frac{m_{\tilde{G}}}{M_2}\simeq 0.04$.   Taking the
string scale $M_{st}=M_P$ and requiring the scale of
{\bf PQ} symmetry breaking $M=10^{12}$ GeV, one gets
$S_3\simeq 1.5\times 10^{15}$ and
$\frac{m_{\tilde{G}}}{M_2}\simeq 0.056$ which is reasonable.}
To conclude, in this model, with $\Lambda=10^5$ GeV, the gravitino, gluino 
and wino masses (at $M$) are given by
\begin{eqnarray}
\label{eq:heavygravitino}
\begin{array}{cc}
m_{\tilde{G}}=18.6\ GeV \\ \\
m_{\tilde{g}}=(\frac{b}{0.1})^2\times 14\ GeV\\ \\
M_2=340\ GeV.
\end{array}
\end{eqnarray}
Hence, either the gluino or the gravitino is the LSP
depending on the particular SUSY breaking model and the value of the parameter $b$.\footnote{In order to have $b = 0.1$ with $M/M_G \sim 10^{-4}$ we need to
take $\lambda^2/\lambda_1 \lambda_2 \sim 10^5$.}

For phenomenological reasons we assume that SUSY
is also broken by the D-term of an anomalous $U(1)_X$ gauge
symmetry as already discussed in Ref. \cite{RabyTobe3}.  Moreover, 
the GMSB and D-term contributions are necessarily comparable.\footnote{ Ref.\ \cite{RabyTobe4} presents a model which dynamically breaks SUSY and
leads to comparable SUSY breaking effects from
gauge-mediated and D-term sources.}
The D-term contribution to scalar masses is given
by
\beq
\label{eq:dtermcont} 
\delta_D\tilde{m}^2_a= d \; Q^X_a \; M_2^2,
\eeq 
where $Q^X_a$ is the $U(1)_X$ charge of the field $a$ and
$d$ is an arbitrary parameter of order $1$ which measures
the strength of D-term versus gauge-mediated SUSY breaking.
The value of $Q^X_a$ for $a = {\bf 16},\;  {\bf 10},\; {\bf 1}$
of SO(10) is given by $1,\; -2, \;4$ \cite{RabyTobe3}.  

We now summarize the messenger scale boundary conditions. The gaugino
masses are given by Eq. \ref{eq:gauginomass} and scalar mass by
Eqs. \ref{eq:scalarmass} and \ref{eq:dtermcont}.\footnote{Note, the
Higgs states also get a SSB mass correction at one loop due to their
direct interaction with the messengers.   This correction is negligible
however because of the small values we have chosen for the free parameters $a, \; b$.}
 D-term SUSY breaking
only contributes to the scalar masses. The SSB trilinear 
scalar coupling $A$ and scalar Higgs mass squared $B$ vanish at tree level
but are generated via RGE running below $M$. 
We have chosen $a=10^{-4}$.  We then determine the free parameters $\Lambda$, $b,\;d$ and $\mu$ (at $M$) and $\alpha_{GUT},\; \epsilon_3$, the top ($\lambda_t$), bottom ($\lambda_b$) and $\tau$ ($\lambda_\tau$) Yukawa couplings (at $M_G$) by fitting the low energy data which we take to include $m_t,\; m_b,\; m_\tau,\; \alpha_{em},\; \alpha_{s}$ and $\sin^2{\theta_W}$.\footnote{ Note, we allow for a small one loop threshold correction to gauge coupling unification at $M_G$ which we have
parametrized by the free parameter $\epsilon_3 = (\alpha_3 - \alpha_G)/\alpha_G$
evaluated at $M_G$.}  Imposing gauge coupling unification at the GUT scale, we renormalize the effective Lagrangian parameters to the EWSB scale using one (two) loop equations for dimensionful (dimensionless) parameters.  
We have also included the effect of a light gluino in the running of
$m_b$ and $\alpha_s$ below the EWSB scale \cite{Clavelli}.  
Finally the one loop SUSY threshold corrections for the $b$ and $\tau$ masses at the EWSB scale have also been included. 

\section{Phenomenology at the EWSB scale}
\label{sec:phenomenology}

The parameter $b$ is varied to keep the gluino pole mass at $30$ GeV. \footnote{Using the RGEs we evaluate
$m_{\tilde g}^{\overline{MS}}$ at $ m_{\tilde g}$ and then calculate the one
loop corrected gluino pole mass.  It is the pole mass which is constrained
to lie between 25 and 35 GeV.  Note, that an 18 GeV $\overline{MS}$ running mass defined at $M_Z$ is roughly equivalent to a 30 GeV pole mass.}   
$d$ and $\Lambda$ set the scale for squark, slepton and gaugino masses.
We have examined cases of fixed $\Lambda$, varying $d$ and vice versa in
order to study the effects of these two SUSY breaking
mechanisms on the low energy phenomenology separately.
$\tan{\beta}$ is solved from Eq. \ref{eq:musquared}
and helps determine the quark and lepton masses. We have 
allowed for values of ($\epsilon_3 < 4\%$). 

In the first part of our analysis, $\Lambda$ is fixed
to the value $\Lambda=10^5$ GeV while $d$ is varied.
Fig.\ \ref{f:Dmufig} shows the values of $|\mu|$ at the messenger
scale giving the best low energy fit. Note that $\mu<0$
and $|\mu|$ increases with $d$. The reason for
negative $\mu$ can best be seen by considering the following equation
\beq
\label{eq:BfromMu}
B=(m^2_{H_u}-m^2_{H_d})\frac{\tan{2\beta}}{2}-
M_Z^2\frac{\sin{2\beta}}{2}.
\eeq
At small $\tan{\beta}$, the second term on the right hand side of the equation
is negligible. Although $m^2_{H_u}$ and $m^2_{H_d}$ have a
common value at the messenger scale, $m^2_{H_u}$ is always
driven to smaller values than $m^2_{H_d}$ by the RGE because
of the larger top Yukawa coupling. We also know that 
$\tan{2\beta}<0$ as long as $\tan{\beta}>1$. We therefore
need $B>0$ at the EWSB scale. The one loop $\beta$-function 
for $B$ is given by
\begin{eqnarray}
\label{eq:Bbetafunc}
\begin{array}{c}
\beta^{(1)}_B=B\{3|Y_t|^2+3|Y_b|^2+|Y_\tau|^2-3g_2^2-
\frac{3}{5}g_1^2\}  \\ \\
+\mu\{6A_tY^\dagger_t+6A_bY^\dagger_b+2A_\tau Y^\dagger_\tau+
6g_2^2M_2+\frac{6}{5}g_1^2M_1\}.
\end{array}
\end{eqnarray}
Since $B$ and the trilinear couplings $A_{t, b, \tau}$ vanish at 
the messenger scale, we must choose $\mu < 0$ in order to get a 
positive $B$ at the EWSB scale. 

The values of $|\mu|, \sqrt{B}$ and $|B/\mu|$ at the $Z$ scale
are also shown in the plot. It is notable that RGE running gives
small values of $B$, eliminating fine tuning in
the Higgs potential, and thus giving a good solution to the
$\mu$ problem.  Note, $|\mu|, \sqrt{B}$ are both increasing functions of $d$;
with a sharper rise at small values of $d$.  This can be understood
using  Eqn. \ref{eq:musquared}.   For moderate values of $d$,  the 
$M_Z^2$ and $m^2_{H_d}$ terms are negligible (the latter due to the
fact $|m^2_{H_d}| < |m^2_{H_u}|$ and the factor of $\tan\beta^2 - 1$
in the denominator).  Hence the approximate relation $\mu^2 = - m^2_{H_u} > 0$
holds.  Increasing $d$ leads to a linear increase in $|m^2_{H_u}|$
(the $U(1)_X$ charge of $H_u$ is -2) and therefore
in $|\mu|$.  For very small values of $d$, $|m^2_{H_u}|$ becomes comparable
to $M_Z^2/2$ and the significant cancellation in the relation
$\mu^2 \approx - M_Z^2/2 - m^2_{H_u}$ leads to a sharp decrease in $\mu$.
Since $B$ is generated from $\mu$ via RGEs, the dependence of $\sqrt{B}$
on $d$ follows that of $\mu$.

We mentioned that $m^2_{H_u}-m^2_{H_d}$ is zero 
at the messenger scale but is negative at the $Z$ scale.
Decreasing $d$ has a small effect on the
value of  $m^2_{H_u}-m^2_{H_d}$ at the $Z$ scale. However from Eq. 
\ref{eq:BfromMu} we see that a sharp decrease in $B$ at small values 
of $d$ has to be compensated by small values of $|\tan{2\beta}|$.
This is the reason for the sharp increase in $\tan{\beta}$,
plotted in Fig.\ \ref{f:Dtanbetafig}, at low values of $d$.\footnote{Note,
a small $|\tan{2\beta}|$ gives a large
$\tan{\beta}$ assuming $\pi/4<\beta<\pi/2$.}

In Fig.\ \ref{f:DHiggsfig} we plot the masses of the Higgs states.
The masses of $A, H^0$ and $H^+$ are determined at tree level
while we have included the one loop SUSY threshold corrections
to the mass of $h$. Since $B$ increases with $d$, the masses of
$A, H^0$ and $H^+$ also increase while $A$ stays the 
lightest. One might think
that since the mass of the lightest Higgs state $h$ is
an increasing function of $\tan{\beta}$, it should decrease
when increasing $d$. However since our model is constrained and
$B$ increases with $d$, the effect of $B$ dominates over
that of $\tan{\beta}$ and the mass of $h$ slowly increases with
$d$. The reason for the slow increase is that the mass of
$h$ is determined by EWSB and hence it cannot be a strong function of $d$.    

In Fig.\ \ref{f:DLHiggsfig}, we have magnified the small $d$ and $\Lambda$ 
regions in order to show the rapid variation in the value of $m_h$ here.
The solid line shows the change in the mass of $h$
versus $d$ for a fixed $\Lambda=10^5$ GeV while the
dashed line shows the variation versus $\Lambda$ with a 
fixed $d$.  As discussed above, $B$ decreases with decreasing $d$.
This decrease in $B$ is directly 
reflected in the decrease in the tree level value of $m_h$.  
Similarly the value of $B$ decreases when $\Lambda$ decreases and thus so
does $m_h$.\footnote{The Higgs mass apparently increases without bound
as $\Lambda$ increases.  We believe this result is due to the fact that we only use
the tree level Higgs potential for EWSB.  Note, we have checked that we
cannot obtain reasonable fits to the data for $\Lambda > 1.1 \times 10^5$ GeV;
hence we believe that our range for the allowed Higgs mass will not be significantly
affected when including higher order corrections to the Higgs potential.}

For completeness, we show the behavior
of $\tan{\beta}$ with changing $\Lambda$ and fixed $d=1$
in Fig.\ \ref{f:Ltanbetafig}; $\tan{\beta}$ decreases rapidly
for $\Lambda<10^5$ GeV. This is because
$|\tan{2\beta}|$ must increase to compensate for $|m^2_{H_u}-m^2_{H_d}|$ 
which is decreasing proportional to $\Lambda^2$ (see Eq. \ref{eq:BfromMu}).  

Fig.\ \ref{f:Dcharginofig} contains plots of the
 masses of charginos and neutralinos versus $d$.
The mass of the heavy chargino ($\chi^+_2$) and the two 
heaviest neutralinos ($\chi^0_3,\chi^0_4$) 
increase, as $|\mu|$ increases with $d$. However, for 
most values of $d$, the mass of $\chi^0_1$ scales as $M_1$ 
while the masses of $\chi^0_2$ and $\chi^+_1$ scale as $M_2$
and do not run with $d$. At very small $d$, $\tan{\beta}$ gets
very large and the off diagonal elements, proportional to
$\sin{\beta}$ in the chargino and neutralino mass matrices, become larger than
the diagonal elements including $\mu$ resulting in a sharp drop in
the masses.   

Squark and Slepton masses are plotted in Fig.\ \ref{f:Dsquarkfig}.
The D-term contributes positively to the masses of squarks and sleptons
since the $U(1)_X$ charges of these fields are $+1$. Consequently,
we see an increase in their masses with increasing $d$. The mixing due to $A$ and $\mu$ terms for the third generation squarks and sleptons
is very small.  Nevertheless, due to the different boundary conditions
at $M$, the right handed squarks and sleptons are lighter than the left handed ones.  The right handed stop  is always the lightest and below $d\simeq 0.7$ it becomes lighter than the top.    
This is because the right handed stop mass squared is driven negative
by RGE as a consequence of the large top Yukawa coupling.

Finally in Fig.\ \ref{f:Lsquarkfig} we plot the masses of squarks,
sleptons, charginos, neutralinos and Higgs fields versus
$\Lambda$. As one expects, all these masses
increase with $\Lambda$ except the lightest Higgs whose
mass is determined by EWSB. 
Recall that all scalar and gaugino masses at the
messenger scale are directly proportional to $\Lambda$.  

\section{Laboratory and cosmological constraints}
\label{sec:experiment}

Let us first consider the heavy gluino.  If it is the LSP it was
shown that it can survive in a narrow window with mass between
25 and 35 GeV \cite{BCG,RabyMafi}. We note that to get this limit, Refs.\ \cite{BCG,RabyMafi} assume very large squark masses.  Ref.\ \cite{RabyMafi} however also argues that lowering the squark masses increases the allowed
range for the gluino mass.   Now consider the possibility that the gravitino
is the LSP and the gluino is the NLSP.  In this case we must check whether the analysis of Refs.\ \cite{BCG,RabyMafi} still applies, i.e. whether
the gluino lifetime is greater than $\sim 10^{-8}\ s$.
The decay rate of the gluino to a gluon and a gravitino is given by

\beq
\Gamma_{\tilde{g}\to g\tilde{G}}=\frac{\alpha_s^2}{48\pi}
\frac{m_{\tilde{g}}^5}{M^2_P m^2_{\tilde{G}}}
(1-\frac{m^2_{\tilde{G}}}{m^2_{\tilde{g}}})^3.
\eeq
Even in the case that $\Lambda_{susy}^2=F_X\simeq 10^{17}\ GeV^2$,
which gives a very light gravitino as in  Eq. \ref{eq:gravitinomass},
a gluino mass of $30$ GeV has a lifetime of
$\tau_{\tilde{g}}\simeq 2\times 10^{6}s\simeq 1$ month. We therefore
conclude that the gluino is in all cases a stable particle with regards to 
detector experiments.  Hence the missing momentum analysis of Refs. 
\cite{BCG,RabyMafi} is relevant and a heavy gluino NLSP with mass in the 
range $25 - 35$ GeV is still viable. 

Note there are several significant advantages for having a heavy gluino LSP
or NLSP. 
\begin{itemize}
\item It reduces the fine tuning necessary for EWSB, since the dominant
contribution to scalar masses due to RG running from $M_G$ to $M_Z$ comes
from color corrections proportional to the gluino mass squared\cite{KaneKing}.
\item Even if the gluino is the NLSP, its lifetime is long enough for 
it to be a candidate for the UHECRon, i.e. the source of the ultra high energy cosmic rays \cite{albuquerque}.
\item  The model with a Higgs mass of order 90 GeV and a stop mass less than the top satisfies some of the dynamical constraints necessary for electroweak baryogenesis in supersymmetric theories \cite{carena}.
\end{itemize}

We now consider the LEP constraints on other SUSY parameters in
our model.  The most important constraints come from the latest Higgs search results at LEP \cite{Cernlatest}. The light neutral 
Higgs $h$ and the CP odd Higgs $A$ in the MSSM are produced 
at LEP via the Higgs-strahlung process $e^+ e^- \to hZ$ 
or the pair production process $e^+ e^- \to hA$. 
$h$ and $A$ decay predominantly into $b\bar{b}$ and 
$\tau^+\tau^-$. Thus LEP experiments search for either a $b\bar{b}$ or       
$\tau^+\tau^-$ plus the decay products of the Z; or
for $bb\bar{b}\bar{b}$ and $\tau^+\tau^- b\bar{b}$.  In our
model the off-diagonal elements in the stop mass-squared matrix are very small, thus the LEP limits for the neutral Higgs in the no stop-quark mixing scenario
are most applicable \cite{Cernlatest,Junk}.  These limits are very
severe.   Looking at our data in
Figs.\ \ref{f:Dtanbetafig} and \ref{f:DLHiggsfig}, it can be seen that
only $d\sim\ 0.40-0.45$ survives the LEP constraint for $\Lambda=10^5$ GeV.
With these values of the parameters, the mass of the lightest neutral Higgs resides in the narrow range $\sim 86 - 91$ GeV with $\tan\beta \sim 9 - 14$. 
At this point, our model also survives the limit on the mass of $A$
as indicated in Ref. \cite{Cernlatest}. $H^+$ and $H^0$ are also too 
massive to be constrained.  For $d\sim\ 0.40-0.45$ and 
$\Lambda=10^5$ GeV we find the lightest stop and neutralino with mass in
the range 100 - 122 GeV and  50 - 72 GeV, respectively.  Note, if $\Lambda$
increases, then the Higgs mass and $\tan\beta$ remain unchanged while all
other masses increase.\footnote{Note, a recent LEP bound on a heavy gluino LSP using stop production and decay \cite{katsenevas} does not constrain the model since the stop mass in our case is larger than the values probed in this search.}

 This narrow region of parameter space is obtained with the fit values of the parameters $\alpha_G = 5.28 - 5.49 \times 10^{16}$ GeV, $\epsilon_3 =  2.44 - 2.42\%,\;\; \lambda_{b, t, \tau} =  (0.065,\ 0.42,\ 0.096) - (0.042,\ 0.42,\ 0.061)$ at $M_G$ and $\mu =  (-85.4) - (-139)$ GeV and $b = 0.06$ at $M$.
Note, we are not able to get a good fit to the data assuming Yukawa coupling
unification.  This was not the case in Ref. \cite{RabyTobe3} and is due to the fact that, unlike the model presented here,  the parameter $B$ was a free parameter.  In the appendix, we show how to complete the model described
earlier in order to accomodate non-unification of Yukawa couplings in this
SO(10) theory.  Clearly this feature of the model is not very satisfying.

The process $e^+\ e^-\to\ $hadrons can constrain the chargino 
mass based on the OPAL 2 $\sigma$ bound on new
physics at $\sqrt{s}$=172 GeV \cite{opal1}. We analysed the contributions
to $e^+\ e^-\to\ $hadrons coming from chargino and neutralino 
pair production using ({\it SPYTHIA, A Supersymmetric Extension 
of PYTHIA 5.7} \cite{spythia}), which has been modified by S. Mrenna 
and K. Tobe to accommodate a gluino LSP. The result is that except for
very small values of $d$ or $\Lambda$ which are strongly ruled out by
the Higgs constraint, the cross section is below the limit
of $3\ pb$ reported in Ref. \cite{opal1}.\footnote{For example, when $\Lambda=10^5$ GeV, the cross section
of the process $e^+\ e^-\to\ $hadrons through charginos and
neutralinos exceeds the 3 pb limit only for $d<0.37$.}

We have evaluated the rate for $b\to s\gamma$.\footnote{In order to do
this calculation we need to assume some values for flavor mixing elements.
As a rough estimate we use the observed CKM matrix elements in the appropriate places.  This is clearly just a rough estimate which gives us at best
an order of magnitude approximation.} The ratio of the SUSY amplitude to the SM amplitude for (varying $d$-fixed $\Lambda$) and (varying $\Lambda$-fixed $d$) are given in Fig.\ \ref{f:DLbsgammafig}. The main
SUSY contribution in our model comes from the charged Higgs-top
loop. Fig.\ \ref{f:DLbsgammafig} shows that we always get an 
amplitude larger than the SM amplitude approaching it at large
$\Lambda$ since $\mu<0$. Note, our result is at the limiting edge
of values allowed by CLEO data \cite{cleo}.  Our result is however incomplete.
We approximate flavor mixing using the known CKM elements as a crude approximation to squark - quark flavor mixing and we only use a one-loop analysis.  With regards to the latter, it is clear that a one loop analysis 
is not sufficient since the higher order
corrections have recently been shown to be very 
important (see for example Ref.\ \cite{Boer}).   Hence within the 
approximations considered here, the model is roughly consistent with
the observed rate for $b \to s\gamma$.

Now consider the low energy consequences of the {\bf PQ} symmetry which
is spontaneously broken at the scale $f_A\equiv M=10^{12}$ GeV. This generates an invisible axion which is a significant component of the energy density of the universe and thus a candidate for cold dark matter \cite{Rosenberg}.  
The axion is predominantly the angular part of the complex scalar field component of the chiral superfield $X$. The radial part of this scalar field and the fermion component of
the supermultiplet are named saxino ($\tilde{S}$)
and axino ($\tilde{A}$) in the literature, respectively.
In our model, these fields obtain large radiative masses
via a messenger ${\bf 10}_A$ loop.

\beq
M_{\tilde{A}}=\frac{\l_X^2}{8\pi^2}\Lambda,
\qquad
M_{\tilde{S}}^2=\frac{\l_X^2}{6\pi^2}\Lambda^2.   
\eeq

The axino decays mainly to a gluino and a gluon.
The decay rate is calculated to be \cite{KmMaNa,Nieves4}

\beq
\label{eq:axinodecay}
\Gamma_{\tilde{A}\to g\tilde{g}}=
\frac{\alpha_{s}^2}{16\pi^3}
f^2(\frac{m_t^2}{\tilde{m}_{t_1}\tilde{m}_{t_2}})
\frac{m_{\tilde{A}}^3}{f_A^2},
\quad
f(x)=\frac{x}{1-x}+\frac{x\ln{x}}{(1-x)^2}.
\eeq
and we find $10^{-22}GeV<
\Gamma_{\tilde{A}\to g \tilde{g}}<10^{-17}GeV$
for $0.2<\frac{m_t^2}{\tilde{m}_{t_1}\tilde{m}_{t_2}}<1$ and
$0.5<\l_X<3$. This translates into an axino lifetime of
$10^{-7}s<\tau_{\tilde{A}\to g \tilde{g}}<
10^{-2}s$.

The saxino decays either to 2 gluons or 2 gluinos.
The decay rate of the saxino to 2 gluons is given
in Ref. \cite{AsaYana}

\beq
\Gamma_{\tilde{S}\to 2g}=\frac{\alpha_s^2}{32\pi^3}
\frac{M^3_{\tilde{S}}}{f_A^2}.
\eeq
For $0.5<\l_X<3$, the decay rate is 
$10^{-18}GeV<\Gamma_{\tilde{S}\to 2g}<
10^{-15}GeV$
with a comparable decay rate into two gluinos.
This gives the saxino an approximate lifetime of
$10^{-9}s<\tau_{\tilde{S}\to 2g}< 10^{-6}s$.
The heavy axino and saxino have very short lifetimes and decay
before nucleosynthesis starts. Their decay therefore does
not affect the standard nucleosynthesis calculations.

Finally, the relic gluino density, assuming the gluino is the LSP, should be small enough that it does not constitute a significant fraction of the dark matter halo density (see for example \cite{starkman}). 
Refs.\ \cite{BCG,Raby1} show that taking into account
the non-purturbative effects of gluino-gluino annihilation
into quark-antiquark and gluon-gluon the relic gluino density is extremely small; $\Omega_{\tilde g} h^2 \sim 10^{-8} - 10^{-11} $.   There are however stringent limits on a gluino LSP coming from searches for anomalous heavy isotopes \cite{MohTep} and from energetic
neutrinos due to gluino annihilations in the sun \cite{faraggi}.  Both of these
latter constraints apparently rule out an absolutely stable gluino LSP.  
They do not however constrain the case of a gluino NLSP and gravitino LSP also
considered in this paper.

\section{Conclusions}

In this paper we have presented a solution to the $\mu$ and strong
CP problems in the presence of a heavy gluino LSP.   The model
has a natural Peccei-Quinn symmetry which prevents the $\mu$ term
at tree level.  However when the {\bf PQ} symmetry is broken at the messenger scale $M\sim 10^{12}$ GeV the $\mu$ term is generated. 

The particle phenomenology of the model is quite novel. 
Either the gluino or the gravitino is the LSP. If the gravitino is the LSP, then the gluino is the NLSP with a lifetime on the order of one month or longer. In either case this heavy gluino, with mass in the range 25 - 35 GeV, can be treated as a stable particle with respect to experiments at high energy accelerators.  

We have studied some of the phenomenological constraints on the model.  
The most significant comes from LEP searches for the neutral Higgs boson.
Our model is most like the no stop-quark mixing benchmark which is severely
constrained by the data.  In fact the model only survives
in a narrow region of parameter space resulting in a light neutral Higgs with mass $\sim 86 - 91$ GeV and $\tan\beta \sim 9 - 14$.  In addition the lightest stop and neutralino have mass $\sim 100 - 122$ GeV and $\sim 50 - 72$ GeV,
respectively.  Thus the model will soon be tested.  Finally, the invisible axion resulting from {\bf PQ} symmetry breaking is a cold dark matter candidate.

\section{ACKNOWLEDGMENTS}
This work was supported by the DOE/ER/01545-796.  A. Mafi thanks K. Tobe for helpful discussions.  S. Raby thanks N. Polonsky and L. Roszkowski for useful
conversations at an early stage in this work.
\section{Appendix}

\appendix

\section{The complete model}
\label{sec:completemodel}

In this section we present the complete model with all the symmetries
and charges. The model is defined at the GUT scale by the SO(10)
invariant superpotential $W\supset W_1+W_2+W_3+W_4+W_5$ 
and a non-renormalizable term in the Kahler
potential $K$ where

\begin{eqnarray}
\begin{array}{cc}
W_1={\bf 16}_3 {\bf 10}_H {\bf 16}_3,\\ \\
W_2=\l_a{\bf 10}_H A {\bf 10}_A+\l_X X {\bf 10}_A^2,\\ \\
W_3=\l_1 \bar{\eta}_1 A\eta_1+
\l_2\bar{\eta}_2 A {\eta}_2+\l X\bar{\bf \eta}_1{\bf \eta}_2,\\ \\
W_4=\l_{Y_3} \bar{\bf \eta}_1 Y {\bf 16}_3,\\ \\
W_5=\l_H {\bf 10}_H \bar{\psi} \bar{\psi}^\prime+
\l_{Y_1} Y \bar{\psi}^\prime \psi^\prime,
\end{array}
\end{eqnarray}

\beq
K\supset \frac{X^\dagger}{M_P}{\bf 10}_H^2+h.c.
\eeq
$({\bf 16}_3, \eta_1, \eta_2, \psi, \psi^\prime)$ are all ${\bf 16}$'s,
$(\bar{\eta}_1, \bar{\eta}_2, \bar{\psi}, \bar{\psi}^\prime)$ are
$\bar{\bf 16}$'s, $({\bf 10}_H,{\bf 10}_A)$ are ${\bf 10}$'s,
$(X,Y)$ are singlets and $(A)$ is an adjoint under SO(10).  $W_1, W_2$ and
$W_3$ were discussed earlier in section \ref{sec:model}.

The theory is invariant under U(1) {\bf PQ} and R symmetries. The charges of
the fields under these symmetries are given in Table \ref{table:charge1}.

\begin{table}[h]
$$
\begin{array}{|c|c|c|c|c|c|c|c|c|c|}
\hline
&&&&&&&&&\\
fields&{\bf 10}_H&{\bf 10}_A&A&X&{\bf \eta}_1&
\bar{\bf \eta}_1&{\bf \eta}_2&\bar{\bf \eta}_2&{\bf 16}_3\\ \hline \hline
R &+1&+1&+2&+2&+1&+1&+1&+1&+3/2\\  \hline
PQ&+1&-1&0&+2&-1/2&+1/2&-5/2&+5/2&-1/2\\  \hline
\end{array}
$$

$$
\begin{array}{|c|c|c|c|c|c|}
\hline
&&&&&\\
fields&\psi&\bar{\psi}&\psi^\prime&\bar{\psi}^\prime&Y\\ \hline\hline
R&0&0&-1/2&+3&+3/2\\  \hline
PQ&0&0&+1&-1&0\\  \hline
\end{array}
$$

\caption{\it The R and {\bf PQ} charges of 
different fields in the complete model.}
\label{table:charge1}
\end{table}

The R symmetry is broken by the vevs of several different fields at the GUT scale. However, the {\bf PQ} symmetry is not
broken at the GUT scale and prevents a $\mu$ term in the superpotential.

$W_4$ contains the field $Y$ which gets a vev 

\beq
\label{eq:vevY}
\langle Y\rangle=M_G. 
\eeq
$W_4$ also results in a bottom-$\tau$ Yukawa coupling non-unification at the
GUT scale as discussed in \ref{sec:nonunif1}.

We use $W_5$ in a standard way to split the top and bottom 
Yukawa couplings by giving vevs to $\psi$ and $\bar{\psi}$
of order $M_G$. This mechanism is discussed 
in \ref{sec:nonunif2}.

\section{Yukawa coupling non-unification in an SO(10) SUSY GUT} 
\label{sec:nonunif}

In minimal SO(10), all standard model
fermions in a given generation are contained in a
single spinor (${\bf 16}$) representation of SO(10). The coupling
of the form $W_1$ 
results in a unified Yukawa couplings for the top and bottom quarks
and the $\tau$ lepton, i.e. $\lambda_t = \lambda_b = \lambda_\tau$ at
$M_G$. It is interesting, and necessary for our model, to see 
if it is possible to relax 
this condition in a simple way. The low $\tan{\beta}$ fit
of our SO(10) SUSY GUT to the infrared-scale 
physical observables requires $\lambda_t- \lambda_b$ splitting at the GUT scale. We are also interested in the possibility of splitting $\lambda_b- \lambda_\tau$. The reason is that our best fits to the data come
from bottom Yukawa couplings which are $\sim 30\%$ smaller
than the $\tau$ Yukawa coupling at the GUT scale. 

\subsection{$\lambda_b- \lambda_\tau$ non-unification}
\label{sec:nonunif1}

$W_1,W_3$ and $W_4$ contain interaction terms

\beq
\label{eq:nonuni1}
\bar{\eta}_1 (\l_1 A \eta_1+\l_{Y_3} Y{\bf 16}_3+\l X \eta_2),
\eeq 
resulting in a heavy (${\bf 16}^{\prime\prime}$) and
a massless (at the GUT scale) (${\bf 16}^{\prime}$) multiplet given by
\beq
{\bf 16}^{\prime\prime}\propto (\l_1 \langle A \rangle \eta_1+\l_{Y_3} \langle Y \rangle {\bf 16}_3)
\eeq
and
\beq
{\bf 16}^{\prime}\propto (\l_{Y3} \langle Y \rangle \eta_1-\l_1 \langle A \rangle {\bf 16}_3).
\eeq
Note, we can safely ignore the last term in Eq. \ref{eq:nonuni1} 
since the vev of $X$ is much smaller than $M_G$.

The massless ${\bf 16}^{\prime}$ is identified with the matter
multiplet containing the third generation quarks and leptons.  
Since $A$ gets a vev in the $B-L$
direction and $B-L$ quantum numbers are different for
quarks and leptons, $\lambda_b- \lambda_\tau$ unification will be
lost, but we still have $\lambda_t - \lambda_b$ unification.
In fact this mechanism is an SO(10) version of the one
introduced in Ref. \cite{DimPom}. Considering
Eqs. \ref{eq:vevA},~\ref{eq:vevY} we define
\beq
s_a=\left(\sin{\theta_{B-L}}\right)_a=  
\frac{\rho T^{B-L}_a}
{\sqrt{1+\rho^2 (T^{B-L}_a)^2} }
\eeq
where $T^{B-L}_a$ is the $B-L$ quantum number of the state
$a=Q,\bar{U},\bar{D},L,\bar{E},\bar{\nu}$ in the matter multiplet
and $\rho=\l_1/\l_{Y_3}$. 
\footnote{In calculating the one loop 
gaugino (two loop scalar) mass contributions from $W_3$ in 
Eq. \ref{eq:gauginomass} (\ref{eq:scalarmass}), we 
have ignored an order
$1/\rho$ correction coming from $W_4$. Since $b$ is a
small free parameter, this does not affect the wino,
bino and scalar masses while the gluino mass is varied
by the free parameter $b$.}   

From $W_1$ we see that the fermions in the
matter multiplet ${\bf 16}^\prime$ obtain mass at the electroweak scale
due to the term
\beq
\left(-s_a {\bf 16}^\prime\right) {\bf 10}_H 
\left(-s_a {\bf 16}^\prime\right). 
\eeq
We therefore have
\beq
\label{eq:lamblamt}
\frac{\l_b}{\l_\tau}=\frac{s_Q s_{\bar{D}}}{s_L s_{\bar{E}}}
=\frac{1}{9} \frac{1+\rho^2}{1+\rho^2/9},
\eeq
where $\l_b$ and $\l_\tau$ are the effective bottom and $\tau$ Yukawa 
couplings.  We thus find
\beq
\frac{1}{9}<\frac{\l_b}{\l_\tau}<1,
\eeq
depending on the choice of $\rho=\l_1/\l_{Y_3}$. For $\rho\simeq 4.2$ 
we get $\frac{\l_\tau-\l_b}{\l_\tau}\simeq 30\%$.

\subsection{$\lambda_t - \lambda_b$ non-unification}
\label{sec:nonunif2}

The splitting between $\lambda_t$ and $\lambda_b,\; \lambda_\tau$ Yukawa couplings is best achieved in the Higgs sector using $W_5$. Since $\psi$
and $\bar{\psi}$ get vevs of order $M_G$,
the Higgs doublet mass term can be written as

\begin{eqnarray}
\left(\begin{array}{ccc}
d_H & d_A& d_{\bar{\psi}^\prime} 
\end{array}\right)
\left(\begin{array}{ccc}
0&0&0 \\
0&\l_X \langle X\rangle&0\\
\l_H \langle\bar{\psi}\rangle&0&\l_{Y_1}\langle Y\rangle
\end{array}\right)
\left(\begin{array}{c}
\bar{d}_H \\ \bar{d}_A\\
\bar{d}_{\psi^\prime}
\end{array}\right).   
\end{eqnarray}
From the above we see that the two light Higgs doublets are $d_H$ and
$\cos{\gamma}\bar{d}_H-\sin{\gamma}\bar{d}_{\psi^\prime}$ which are identified
with the light Higgs doublets of the MSSM, $(H_u, H_d)$, respectively. 
$\cos{\gamma}$ is given by
\beq
\cos{\gamma}=\frac{\rho^\prime}{\sqrt{1+\rho^{\prime 2}}}
\eeq
where
\beq
\rho^\prime=\frac{\l_{Y_{1}}\langle Y\rangle}
{\l_H \langle\bar{\psi}\rangle}.
\eeq
Note, the rest of the Higgs doublets remain very heavy.
From $W_1$ we see that
\beq
\frac{\l_b}{\l_t}=\cos{\gamma}
\eeq
where $\l_t$ and $\l_b$ are the effective Yukawa couplings of the
top and bottom quarks. A hierarchy of 50 is easily achieved by
choosing $\rho^\prime\simeq 1/50$. 

\section{$\tau$ neutrino mass}
\label{sec:neutrinomass}

In this section we show that it is possible to get a reasonable $\tau$
neutrino mass, in agreement with atmospheric neutrino oscillations.
Let us add two SO(10) singlets $N$ and $P$ to the model with R and {\bf PQ} charges
given in Table \ref{table:charge2}.

\begin{table}[h]
$$
\begin{array}{|c|c|c|}
\hline
&&\\
fields&$N$&$P$\\\hline\hline
$R$&+5/2&-1\\  \hline 
PQ&+1/2&-1\\  \hline
\end{array}
$$

\caption{\it The R and {\bf PQ} charges of
the new fields in the neutrino sector.}
\label{table:charge2}
\end{table}
The only possible couplings for these singlets are
\beq
\l_{n_1} \bar{\psi} N {\bf 16}_3+\l_{n_2}N^2P.
\eeq
We assume that $P$ gets a vev of order the
messenger scale ($10^{12}$ GeV). The neutrino mass matrix is then given by

\begin{eqnarray}
\left(\begin{array}{ccc}
\nu_L&\bar{\nu}_R&N
\end{array}\right)
\left(\begin{array}{ccc}
0&m_t \frac{s_L s_{\bar{E}}}{s_Q s_{\bar{U}}}& 0\\
m_t \frac{s_L s_{\bar{E}}}{s_Q s_{\bar{U}}}&0&
-s_{\bar{E}}\l_{n_1}\langle \bar{\psi}\rangle \\
0&-s_{\bar{E}}\l_{n_1}\langle \bar{\psi}\rangle&\l_{n_2}\langle P\rangle
\end{array}\right)
\left(\begin{array}{c}
\nu_L\\ \bar{\nu}_R \\N
\end{array}\right).
\end{eqnarray}
The mass of the lightest neutrino, identified as $\nu_\tau$, is given by

\beq
\label{eq:neutrinomass}
m_{\nu_\tau}=(\frac{s_L}{s_Q s_{\bar{U}}})^2 \frac{\l_{n_2}}{\l_{n_1}^2}
\frac{m_t^2\langle P\rangle}{\langle \bar{\psi} \rangle^2}\simeq
6.3\times 10^{-7} (\frac{\l_{n_2}}{\l_{n_1}^2})
(\frac{10^{16}GeV}{\langle \bar{\psi} \rangle})^2 \; {\rm eV}  \sim 6 \times 10^{-2}\; {\rm eV}.
\eeq
Thus one gets a reasonable value for $ m_{\nu_\tau}$ with $\l_{n_1} \sim 3 \times 10^{-3}$ and $\l_{n_2} \sim 1$.

\begin{center}
\begin{figure}
\scalebox{1.0}[1.0]
{
\includegraphics{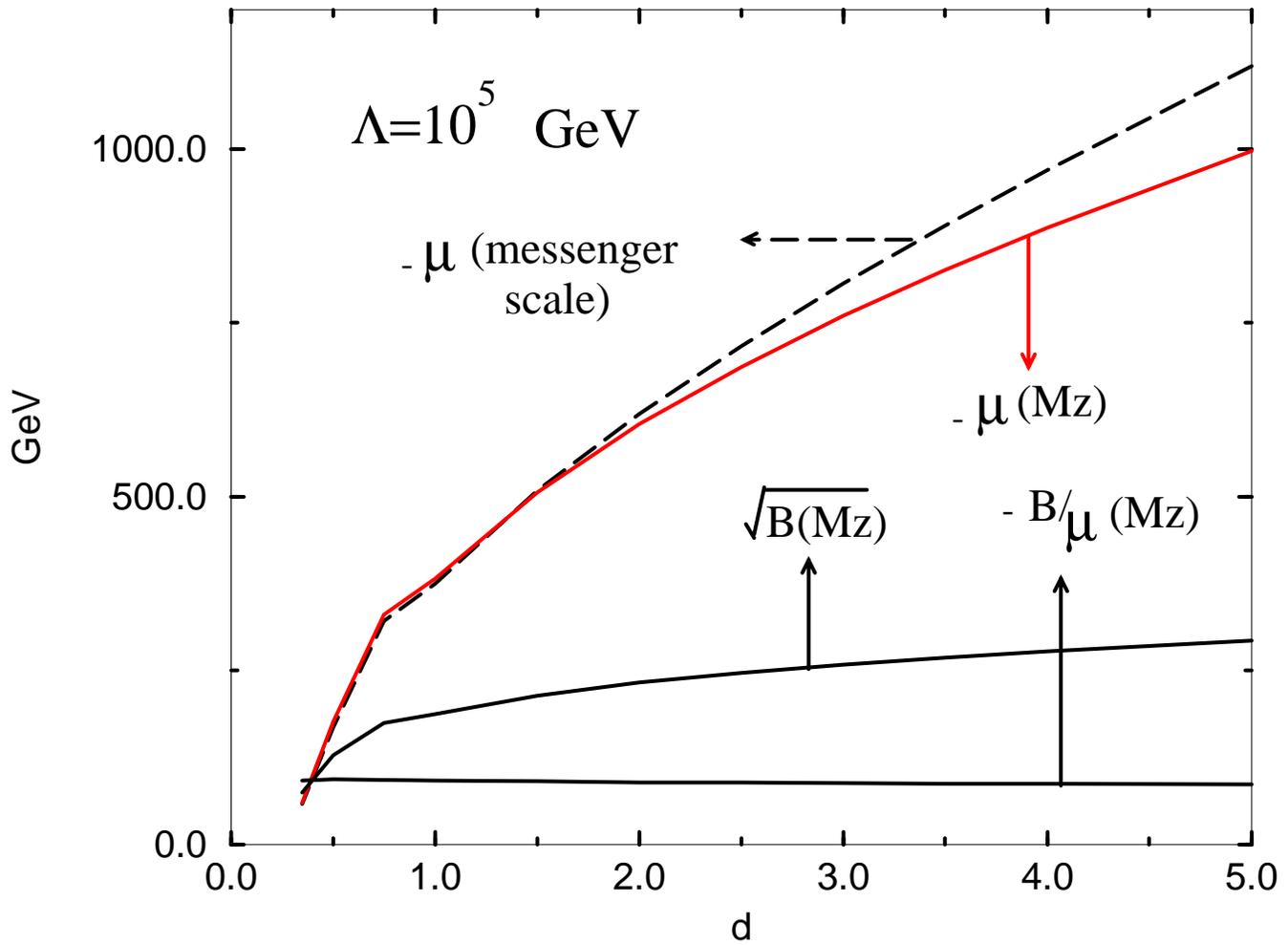}
}
\caption{$-\mu$ at the messenger scale, $-\mu,-B/\mu$ and 
$\sqrt{B}$ at the EWSB scale are plotted in this figure.}
\label{f:Dmufig}
\end{figure}

\begin{figure}
\scalebox{1.0}[1.0]
{
\includegraphics{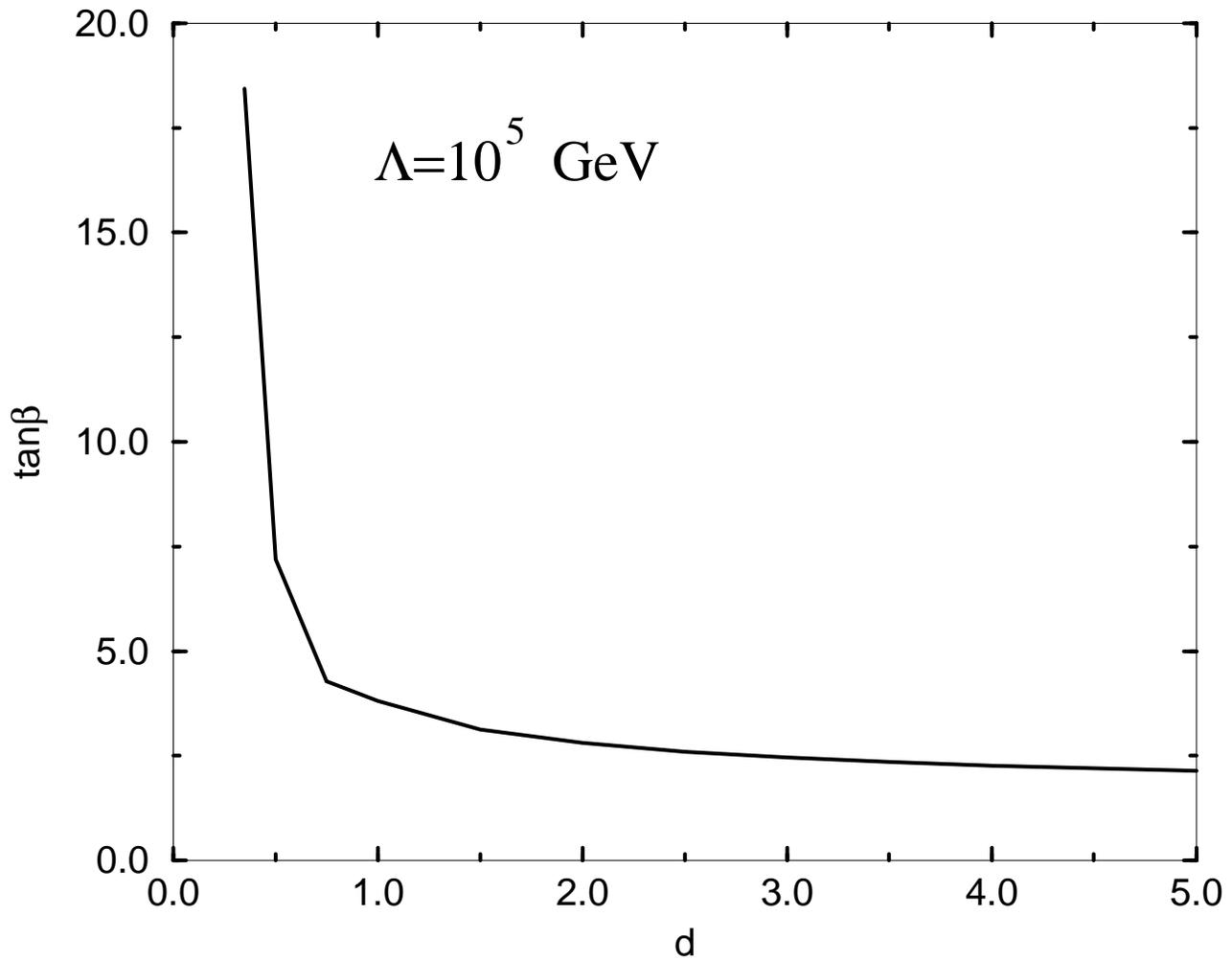} 
}
\caption{Variation of $\tan{\beta}$ versus $d$ for fixed $\Lambda=10^{5}$ GeV.}   
\label{f:Dtanbetafig} 
\end{figure}

\begin{figure}
\scalebox{1.0}[1.0]
{
\includegraphics{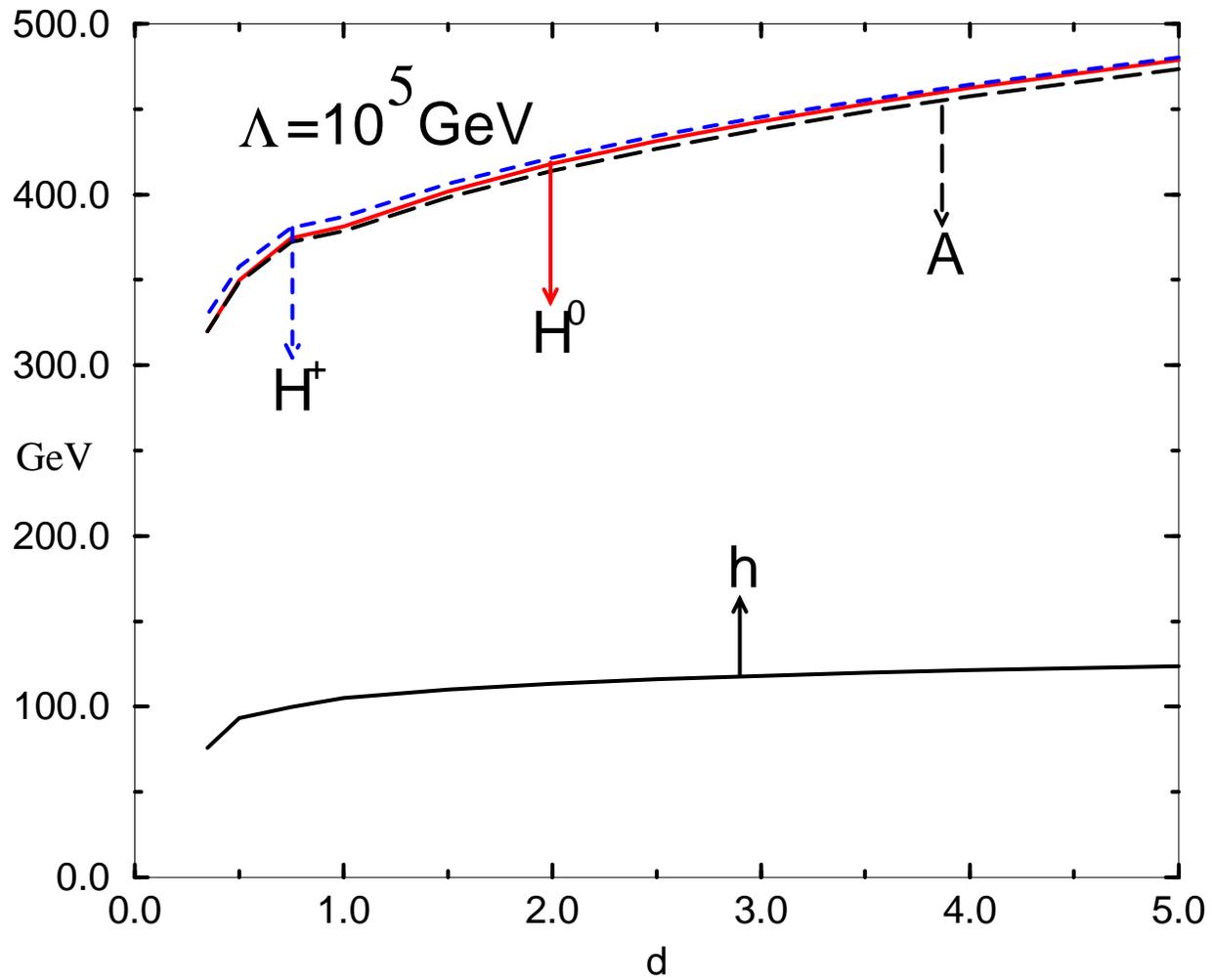} 
}
\caption{MSSM Higgs masses                 
versus $d$ for fixed $\Lambda=10^{5}$ GeV.}
\label{f:DHiggsfig} 
\end{figure}

\begin{figure}
\scalebox{1.0}[1.0]
{
\includegraphics{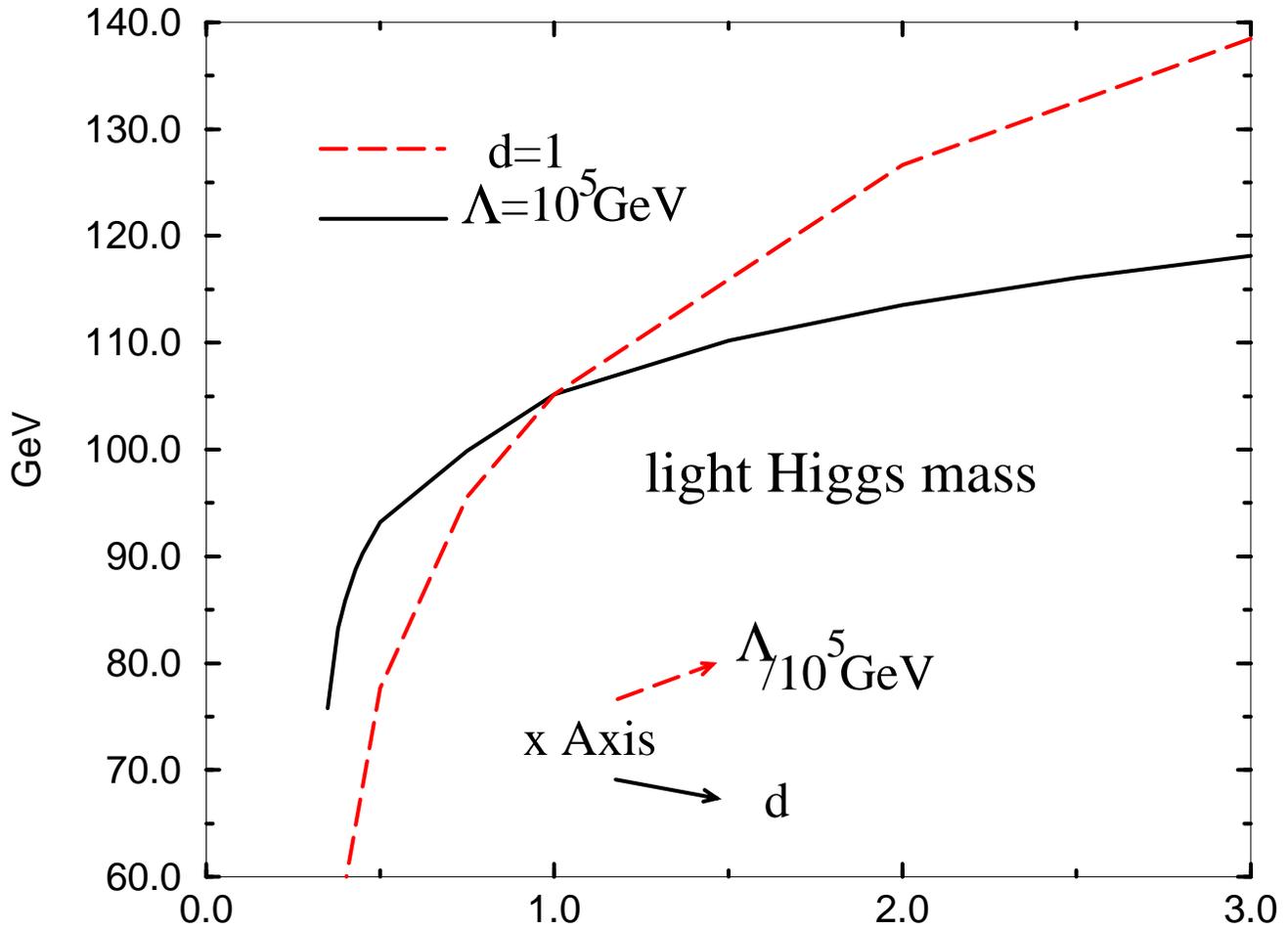}
}
\caption{A magnified variation of the mass of the lightest neutral Higgs, $h$
for small $d$ and $\Lambda$. The solid line shows the variation
versus $d$ for a fixed $\Lambda=10^5$ GeV while the
dashed line shows the variation versus $\Lambda$ with a
fixed $d=1$.
}
\label{f:DLHiggsfig}
\end{figure}

\begin{figure}
\scalebox{1.0}[1.0]
{
\includegraphics{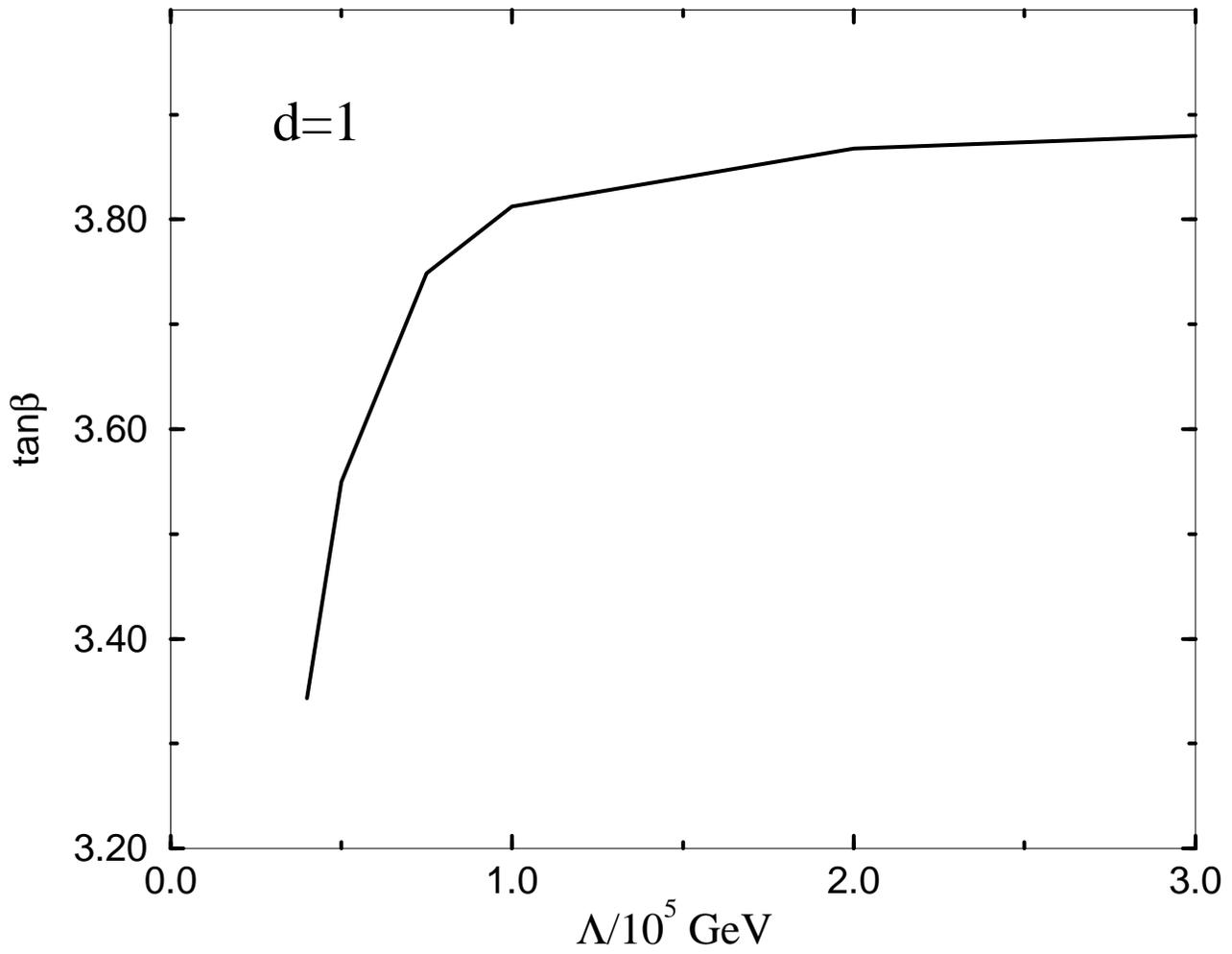}
}
\caption{Variation of $\tan{\beta}$ versus $\Lambda$ for fixed $d=1$.}
\label{f:Ltanbetafig}
\end{figure}

\begin{figure}
\scalebox{1.0}[1.0]
{
\includegraphics{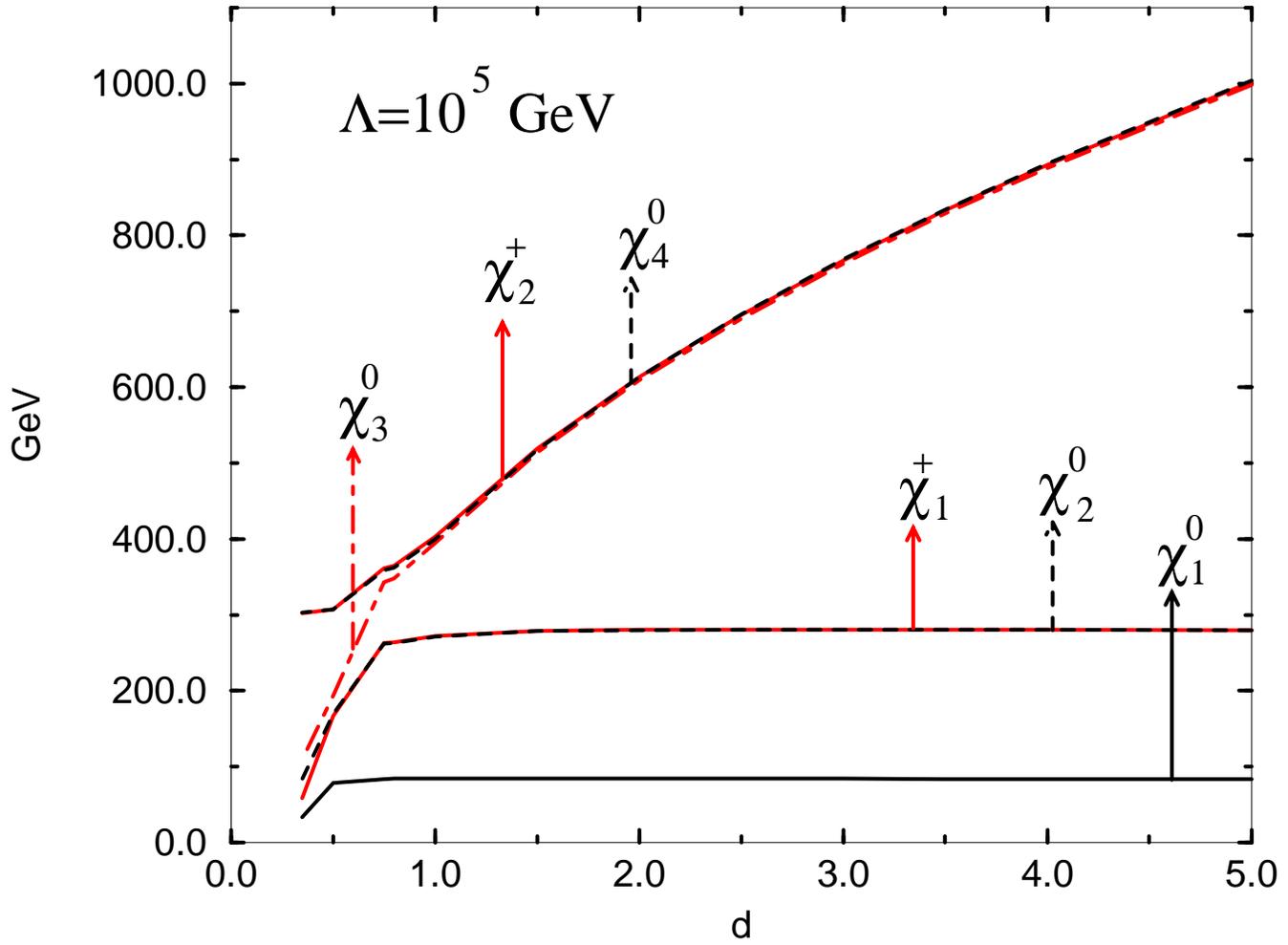}
}
\caption{The Chargino and neutralino masses 
versus $d$ for fixed $\Lambda=10^{5}$ GeV.}
\label{f:Dcharginofig}
\end{figure}

\begin{figure}
\scalebox{1.0}[1.0]
{
\includegraphics{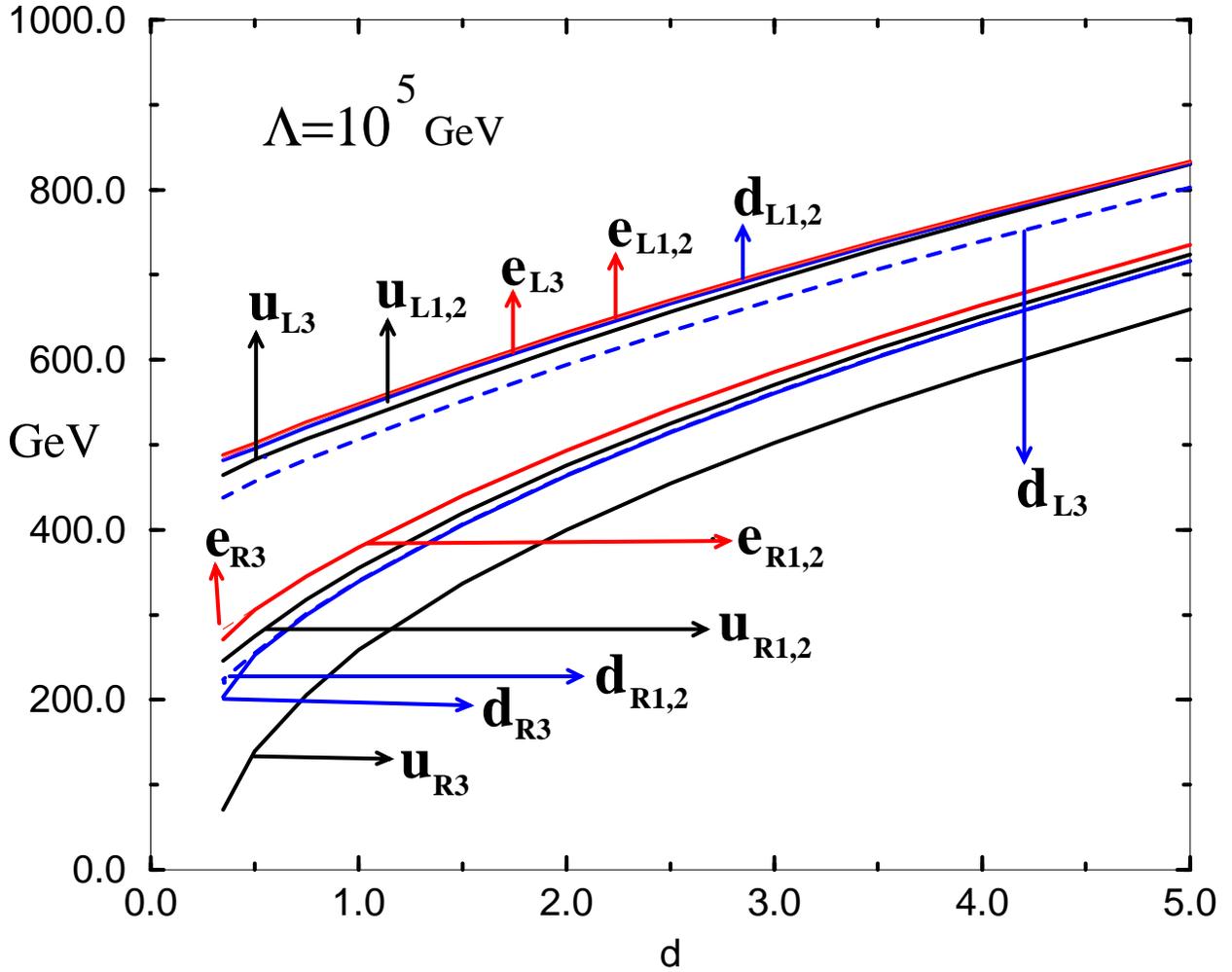}
}
\caption{Squark and slepton masses                 
versus $d$ for fixed $\Lambda=10^{5}$ GeV.}
\label{f:Dsquarkfig}
\end{figure}

\begin{figure}
\scalebox{1.0}[1.0]
{
\includegraphics{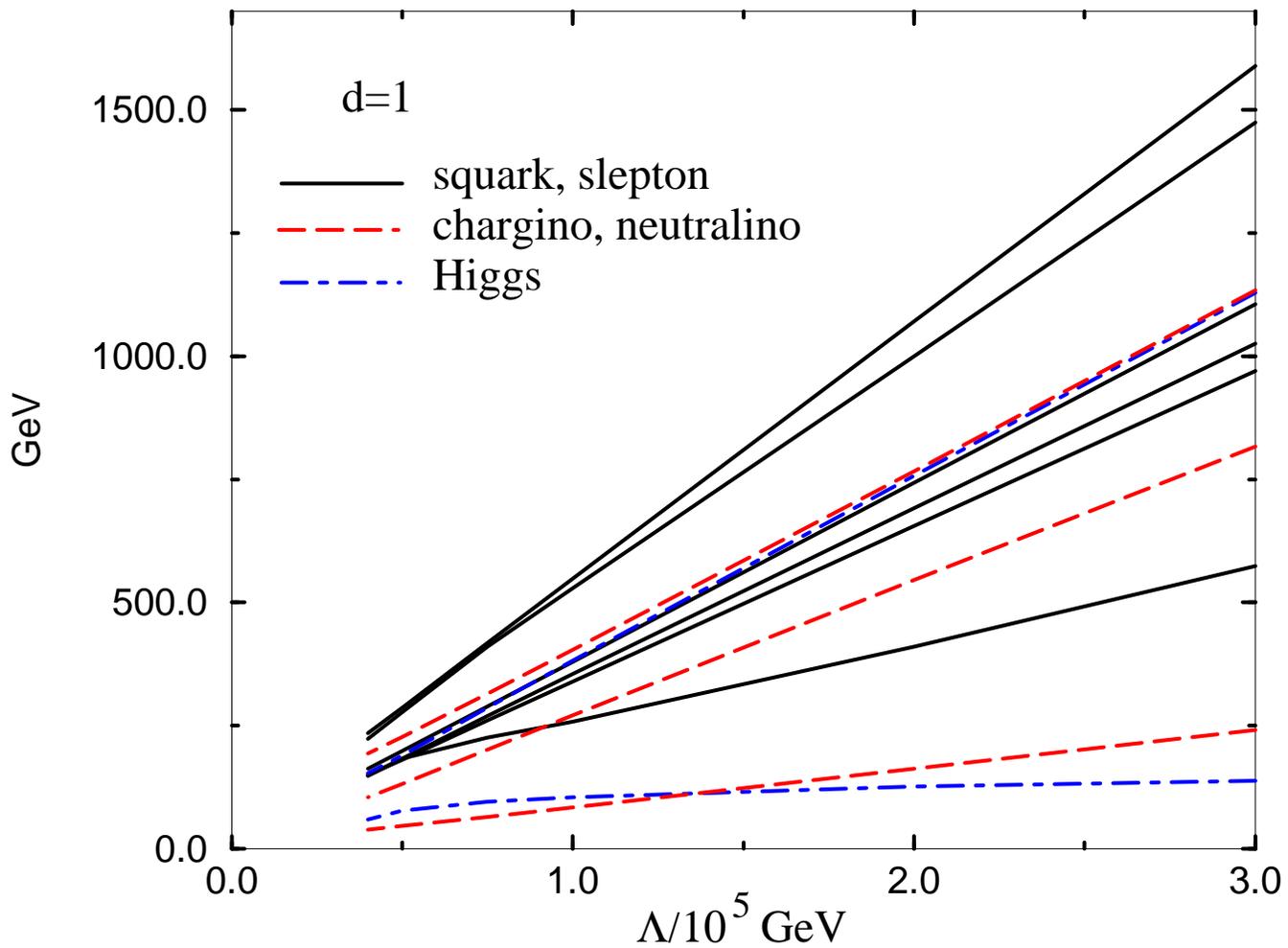}
}
\caption{Squark, slepton, chargino, neutralino and Higgs masses
versus $\Lambda$ for fixed $d=1$. Many of the masses are almost
degenerate, therefore one representative is shown from each set.}
\label{f:Lsquarkfig}
\end{figure}

\begin{figure}
\scalebox{1.0}[1.0]
{
\includegraphics{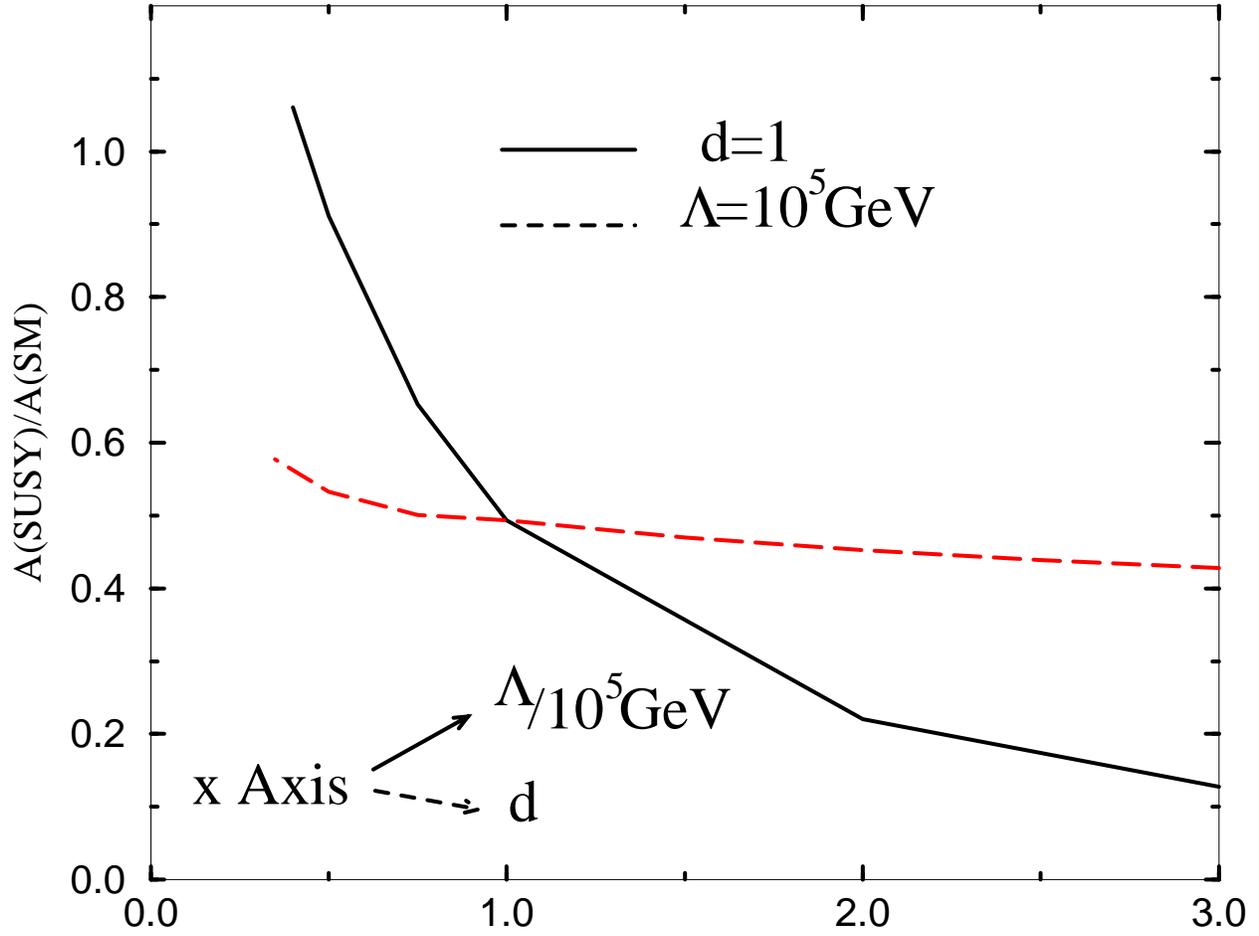}
}
\caption{The ratio of the SUSY contribution to the amplitude of
$b\to s\gamma$ to the SM contribution. The dashed line shows the variation
versus $d$ for a fixed $\Lambda=10^5$ GeV while the
solid line shows the variation versus $\Lambda$ with a
fixed $d=1$.}
\label{f:DLbsgammafig}
\end{figure}
\end{center}

\end{document}